\documentclass[superscriptaddress,nofootinbib]{revtex4}
\usepackage{mathrsfs}
\usepackage{verbatim}
\usepackage{amsmath}
\usepackage{graphicx}
\usepackage{color}
\definecolor{darkgreen}{rgb}{0,0.5,0}
\definecolor{violet}{rgb}{0.5,0.5,1}

\newcommand{\be}{\begin{equation}}
\newcommand{\ee}{\end{equation}}
\newcommand{\bea}{\begin{eqnarray}}
\newcommand{\eea}{\end{eqnarray}}
\newcommand{\bi}{\begin{itemize}}
\newcommand{\ei}{\end{itemize}}

\newcommand{\A}{\psi_A}
\newcommand{\Abar}{\overline{\psi}_A}
\newcommand{\mA}{m_A}
\newcommand{\B}{\psi_B}
\newcommand{\Bbar}{\overline{\psi}_B}
\newcommand{\mB}{m_B}

\begin{document}

\begin{flushright}
SLAC-PUB-16117\\
UMD-PP-014-016\\
SU-ITP-14/24
\end{flushright}
\title{Detecting Boosted Dark Matter from the Sun with Large Volume Neutrino Detectors}
\author{Joshua Berger}
\affiliation{SLAC National Accelerator Laboratory, Menlo Park, CA
  94025, USA}
\author{Yanou Cui}
\affiliation{Perimeter Institute for Theoretical Physics, Waterloo, Ontario N2L 2Y5, Canada }
\affiliation{Maryland Center for Fundamental Physics, University of Maryland, College Park, MD 20742, USA}
\author{Yue Zhao}
\affiliation{Stanford Institute of Theoretical Physics, Physics Department,\\
Stanford University, Stanford, CA 94305, USA}

\begin{abstract}

We study novel scenarios where thermal dark matter (DM) can be efficiently captured in the Sun and annihilate into boosted dark matter. In models with semi-annihilating
DM, where DM has a non-minimal stabilization symmetry, or in models with
a multi-component DM sector, annihilations of DM can give rise to
stable dark sector particles with moderate Lorentz boosts.  We investigate both of these
possibilities, presenting concrete models as proofs of concept.
Both scenarios can yield viable thermal relic DM with masses
O(1)-O(100) GeV. Taking advantage of the energetic
proton recoils that arise when the boosted DM scatters off matter, we
propose a detection strategy which uses large volume terrestrial detectors, such
as those designed to detect neutrinos or proton decays.  In particular, we propose a search for  
proton tracks pointing towards the Sun.  We focus on signals at
Cherenkov-radiation-based detectors such as Super-Kamiokande (SK) and
its upgrade Hyper-Kamiokande (HK). We find that with spin-dependent
scattering as the dominant DM-nucleus interaction at low energies,
boosted DM can leave detectable signals at SK or HK, with sensitivity comparable to DM direct detection experiments while being consistent with current constraints. 
Our study provides a new search path for DM sectors with non-minimal structure.
\end{abstract}

\preprint{SLAC-PUB-16117}
\preprint{UMD-PP-014-016}
\preprint{SU-ITP-14/24}
\email{ jberger@slac.stanford.edu, ycui@perimeterinstitute.ca, zhaoyue@stanford.edu}

\maketitle

\tableofcontents

\section{Introduction}
The evidence for the existence of particulate Dark Matter (DM)
\cite{Zwicky:1933gu,Begeman:1991iy,Bertone:2010zza} is extremely
compelling.  All of the robust signals seen thus far are, however,
gravitational and do not pin down the detailed properties of DM.  In
order to proceed, it is imperative to discover possible non-gravitational interactions of DM with
other particles.  One discovery strategy is to look for
scattering of ambient DM off of SM nuclear targets. This is well
motivated by the paradigm of Weakly Interacting Massive  
 Particle (WIMP) DM, where thermal annihilation to SM states
 predicts DM relic abundance, and generally implies  
 an appreciable DM-nucleon scattering rate based on crossing-symmetry.
The majority of halo DM today is expected to be deeply non-relativistic, having a
typical velocity of order $10^{-3}$.  The momentum transfer in
collisions of such DM with hadronic matter is only of order a few keV,
requiring specialized instruments with low thresholds for the
detection of recoils.  Such experiments are powerful probes of DM
interactions, particularly in scenarios where DM sector is minimal and has
s-wave, Spin Independent (SI) interactions with nuclei, which are
enhanced due to the coherence of the interactions over the entire
nucleus.  In non-minimal scenarios, especially when the SI interaction
is insignificant, other detection strategies could 
offer significantly enhanced sensitivity to DM interactions.

One compelling possibility is to look for ``boosted DM'' that is not of thermal, 
cosmological origin, but rather is generated at late times, as
introduced in \cite{Huang:2013xfa,Agashe:2014yua,Detmold:2014qqa}.  
Several well-motivated classes of models, such as multi-component DM
models and models with non-minimal stabilization symmetries, can
generate boosted DM as a product of annihilation or decay in nearby clumps
of DM.  The relevant processes have forms such as multi-component
annihilation $\psi_i \psi_j \to \psi_k \psi_\ell$ 
\cite{DEramo:2010ep, SungCheon:2008ts, Belanger:2011ww},
semi-annihilation $\psi_i \psi_j \to \psi_k \phi$ (where $\phi$ is a
non-DM state)
\cite{DEramo:2010ep,Hambye:2008bq,Hambye:2009fg,Arina:2009uq,Belanger:2012vp}, 
$3 \rightarrow 2$ self-annihilation
\cite{Carlson:1992fn,deLaix:1995vi,Hochberg:2014dra}, or decay
transition $\psi_i \to \psi_j + \phi$.  Annihilation of DM is of
particular interest, as it is required in the generic scenario that
the DM abundance is set by thermal freeze-out.

In this paper, we focus on two concrete scenarios: models with
semi-annihilation of one DM species $\psi$ charged under a $Z_3$
symmetry, and models with a two-component DM sector with species $\A$
and $\B$ having masses $m_A > m_B$ and $\A$ being the dominant DM
component (here, $\psi$ and $\psi_{A,B}$ need not be fermions).  In
the example of semi-annihilation of $Z_3$ DM, the DM thermal relic
abundance is set by the annihilation process
\be
\psi \psi\rightarrow \bar{\psi}\phi,
\ee
where $\phi$ is a lighter dark sector state that may decay away. At
present times, the non-relativistic $\psi$ undergoes the same
annihilation process in the galactic halo or in the Sun if it is
captured there. Assuming $m_\phi\ll m_\psi$, the final state
$\bar{\psi}$ is produced with a Lorentz boost factor
$\gamma=5/4$.  In the example of two-component DM sector,
thermal relic abundance of dominant DM $\A$ is set by:
\be
\label{eq:AAtoBB}
\A \Abar \to \B \Bbar.
\ee
The same annihilation in the present day produces final state $\B$
with Lorentz factor $\gamma = \mA / \mB$.

In the earlier work \cite{Agashe:2014yua} on boosted DM detection, one 
of us (YC) and collaborators focused on the above two-component DM
model as an example, assuming $\A$ has no direct couplings to
SM.  In the particular
scenario considered in that work, $\B$ had interactions with both
electrons and quarks, making electrons the more sensitive
scattering target for detection provided that the mediator of these
interactions was sufficiently light.  If interactions with electrons are
suppressed or if the mediator is heavier than $O(10~{\rm MeV})$, then
the detection via interactions with quarks is important.  These
interactions can also enhance the flux of boosted DM by generating
a large rate of DM capture and subsequent annihilation in the Sun.
For a broad range of parameters, the flux from the Sun will dominate
over the flux from the galactic center, making it possible to have
observable signals with scattering cross  sections of weak scale size
or even smaller.  Both DM and mediator masses could occupy a wide
range of $O(1-100)$ GeV. This paper therefore  
addresses the scattering off of hadrons as a means of detection of
boosted DM.  

The large boost factor of this flux of DM opens new avenues for
its detection via hadronic recoil.  The low threshold requirement for
detection of non-relativistic DM is relaxed and it 
becomes more effective to use much larger detectors that are sensitive
only to more energetic recoils.  The boost factors obtained in
semi-annihilating and two-component DM models are typically
non-negligible, but modest, as seen above.  The typical momentum
transfer is $O({\rm GeV})$ for boosted DM masses larger than a few
GeV.   The largest experiment of sensitive to hadronic recoils at this
energy is currently Super-Kamiokande (SK) \cite{Fukuda:2002uc}.
Several planned and proposed experiments could have sensitivity,
including not only Hyper-Kamiokande (HK) \cite{Abe:2011ts}, but liquid
Argon based detectors such as l LAr TPC and GLACIER
\cite{Bueno:2007um,Badertscher:2010sy}.  Other large volume detectors
such as IceCube/PINGU/MICA \cite{Ahrens:2002dv,Aartsen:2014oha, MICA},
KM3NeT \cite{Katz:2006wv}, and ANTARES \cite{Collaboration:2011nsa}
are best suited to looking for higher energy recoils.  Note that these experiments
were designed to look for neutrinos and/or proton decay, but can
be repurposed to find boosted DM.  This is not surprising as
interactions of boosted DM with hadronic matter are similar in
structure to neutral-current interactions of high energy neutrinos.

\begin{figure}
  \centering
  \includegraphics[width=0.9\textwidth]{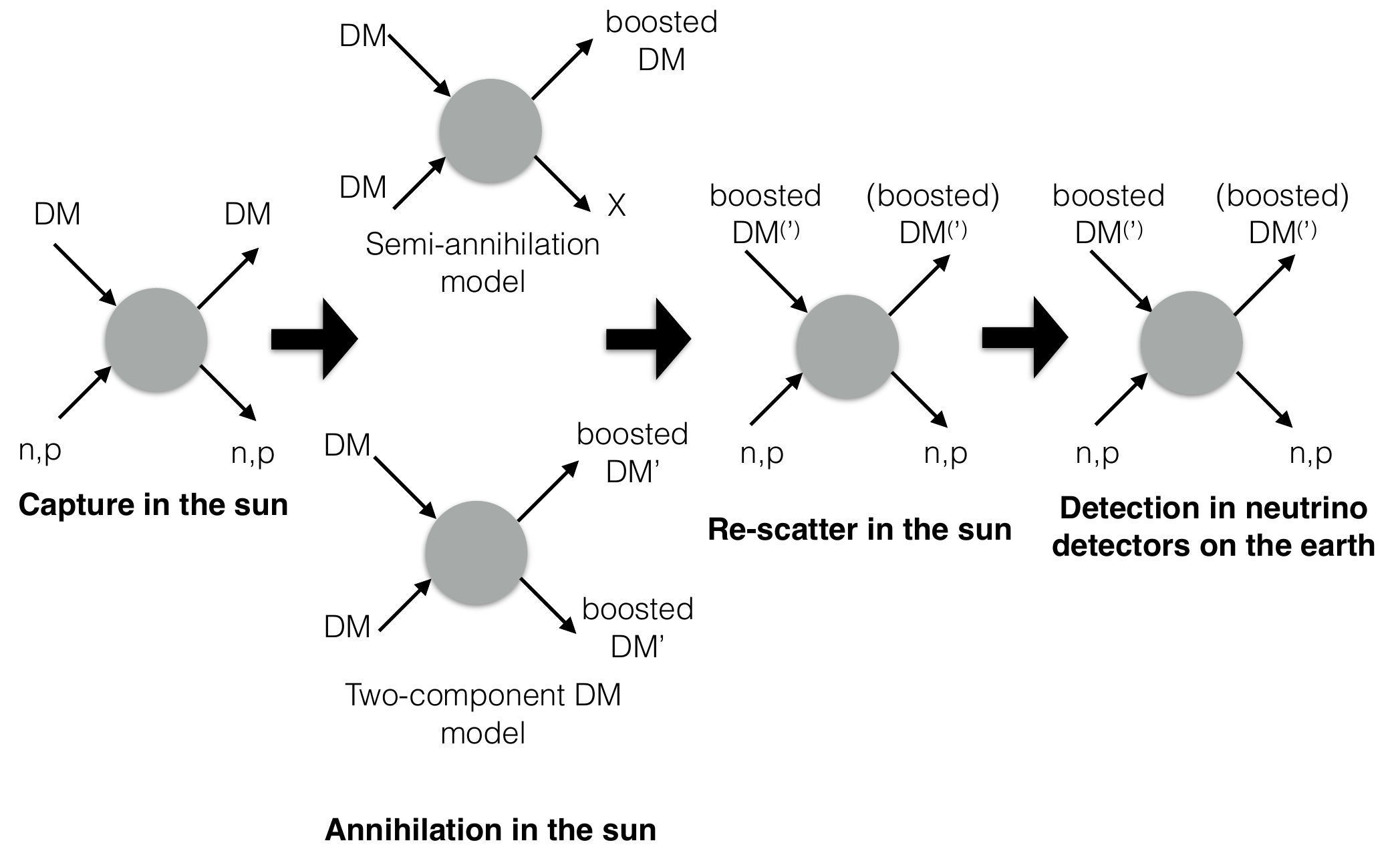}
  \caption{The chain of processes leading to boosted DM signal from
    the Sun. DM' denotes the lighter DM in the two-component DM
    model. $X$ is a lighter dark sector particle that may decay
    away. Details of the two example models, semi-annihilating DM and
    two-component DM, are given in Section \ref{sec:models}.}
    \label{fig:processes}
\end{figure}

Compared to the boosted DM signal from DM annihilation in
the GC as studied in \cite{Agashe:2014yua}, the signal from the Sun
involves more processes, including the capture,
evaporation, annihilation, and rescatter (slow-down) of the DM
particles in the Sun, as well as the scattering of boosted DM off
nuclei in the terrestrial detectors. In addition, there is the
apparent challenge that an appreciable DM-nucleon scattering rate
giving rise to detectable signals is likely to be ruled out by the
existing DM DD bounds on the thermal non-relativistic
component of DM. In our work, we have taken all the above complexities
into consideration, and found that there are large classes of models
with reasonable range of parameter space giving good detection
prospect at SK and HK, while being compatible with conventional DM
detection limits. The key feature of these viable models is that the
DM scattering off nuclei is dominantly Spin-Dependent (SD) and/or has a
velocity-dependence, such as $v^2$ in the non-relativistic limit. We also
note that the the signal reach can be further improved by using
ionization-based liquid argon neutrino detectors where Cherenkov
threshold is irrelevant.   In Fig.~\ref{fig:processes} we illustrate the
chain of processes involved in giving rise to the boosted DM signal in
which we are interested.

A search for an excess of proton recoils pointing toward the Sun as we propose here is a
new search that has not been performed by the SK collaboration.  In
the search for neutrinos from WIMP DM
annihilation in the Sun, electron or muon tracks along the direction
of the Sun, dominantly from charged-current interaction have been
studied \cite{Desai:2004pq,Mijakowski:2012dva,Mijakowski:slides},
while similar proton recoils from neutral-current (NC)
interaction were not investigated due to the relatively larger scattering angle. Nonetheless, as
discussed earlier, boosted DM only scatter off SM via NC-type of
interactions. Unless very light mediator is involved,
a proton recoil signal would typically have much larger rate compared to
a single-ring electron signal and is the primary search channel.

We would also like to reiterate other broad motivations for
investigating boosted DM signals, as have been discussed in more
detail in \cite{Agashe:2014yua}. 
First, annihilation into dark sector particles, as occurs in both our
two-component DM or semi-annihilating DM examples, evades
strong constraints from DM detection experiments in a natural way,
while still allowing for a thermal freeze-out origin of DM. This is complementary to other
variations such as models where 
DM annihilates into dark radiation or to dark states that decay back to
the SM, as discussed in Ref.\ 
\cite{Pospelov:2007mp, ArkaniHamed:2008qn, Ackerman:mha,
  Nomura:2008ru, Mardon:2009gw, AB_CMB} for example.
Second, these studies of boosted DM demonstrate how the expected
phenomenology, and search strategies for a non-minimal DM
sector or single component DM with non-minimal stabilization symmetry
can be very different from those for conventional single-component
$Z_2$ DM models. Non-minimality is already a familiar fact for the SM
matter sector, where protons and electrons stabilized by separate $B$-
and 
$L$-number symmetries. It is quite reasonable that DM sector also has beyond the simplest, minimal content. Earlier work on phenomenology from multi-component DM sector include 
\cite{Hodges:1993yb, Mohapatra:2000qx,Berezhiani:2000gw,
Foot:2001hc,Fairbairn:2008fb,Zurek:2008qg,Profumo:2009tb, Fan:2013yva}, on DM sector with non-minimal stabilization symmetry
such as $Z_3$ include \cite{D'Eramo:2012rr,Boehm:2014bia,Aoki:2014cja,Ko:2014loa,Aoki:2014lha}.

The organization of this paper is as follows. In
Sec.~\ref{sec:models} we present examples of semi-annihilating DM
and two-component DM models in detail, starting from effective
operators for DM interactions with quarks and then developing UV
completions. In Sec.~\ref{sec:flux} we analyze the various DM
processes in the Sun including capture, evaporation, annihilation
and rescattering, and eventually determine the flux of boosted DM
incident on the Earth. Then in Sec.~\ref{sec:detect} we discuss the
detection rate of boosted DM at large volume detectors using the
proton Cherenkov ring signal.  We assess the discovery prospects at
SK and its upgrade HK in Sec.~\ref{sec:results}, commenting on
relevant constraints on these particular models. Finally we conclude
in Sec.~\ref{sec:conclusions} with discussions of other
possibilities. We include more details in the appendices.

\section{Models}\label{sec:models}
In this section, we present the concrete examples of the two classes
of models on which we focus in this paper, semi-annihilating DM
models and two-component DM model.  These models are related by the
fact that DM particles appear in the final state of annihilation
processes, opening the possibility that the final state DM is
boosted.

 In the most concise version of a semi-annihilation model,
there is only one specie of dark matter particle.  There is thus a
direct relation between solar capture rate and the detection rate at
neutrino detectors such as SK. Furthermore, there is only a
small yet generic range ($\gamma=1-1.25$) for the boost factor of
final state DM particle in the semi-annihilation models, which as we
will see, interestingly falls in the sweet-spot for detecting a proton
track signal.

In the two component DM models, different particles are involved in solar capture and SK
detection processes, leading to more general kinematic possibilities
at the expense of a larger number of parameters.  The mass ratio
between two DM particles in the two component DM model controls the
boost factor of DM particle in the final states, which will be
important in determining the signatures at SK and other detectors. We
demonstrate below that the preferred mass ratio range for obtaining an
observable signal ranges approximately from 1.1 to 2.2. If the mass
ratio is too low, then the DM particle in the final state is not
boosted and the recoiling proton does not generate Cherenkov light. If
the ratio is too high, the interaction between the boosted DM and the
protons in the detector is dominantly inelastic and the rate of single
Cherenkov ring events, which are most easily distinguished from background
events, is suppressed.

In studying these signatures, however, we parameterize both classes
of models in terms of a small number of phenomenological parameters
which are relatively insensitive to the details of the complete
models we present in this section.  These models serve as an
important proof-of-principle that complete models can be
constructed, as well as a motivation for the forms of interactions
that we will introduce.  Other models yielding similar signatures
may be possible, but the study of these other possibilities is
beyond the scope of this work.

\subsection{Semi-annihilating DM models}

There are several simple ways to construct a semi-annihilating DM
model. For example, introducing either a $Z_3$ symmetry for DM
particles or a near-degenerate spectrum in the dark sector along with a
stabilizing symmetry can lead to such behavior.  In this paper, we
focus only on the simplest version of semi-annihilation model where we assume a single DM particle $\chi$
charged under a $Z_3$ symmetry, which protects $\chi$ from decay.  The
DM $\chi$ can either be a scalar or a fermion. Taking the
approximation that DM particles in the initial states are almost at
rest, the boost factor of DM particle in the final state is
\begin{equation}
  \label{eq:BoostFactor}
  \gamma_{\chi}=\frac{(5m_{\chi}^2-m_{\psi}^2)}{4m_\chi^2},
\end{equation}
where $m_{\psi}$ is the mass of the lighter unstable particle $\psi$ in the
final state of semi-annihilation process, $\psi$ can either be a
scalar or fermion. $\gamma_\chi$ ranges from
1 to 1.25. For simplicity, we assume $m_{\psi}\ll m_\chi$. Then
the boost factor is near maximal. Lowering $\gamma_\chi$ by a small
amount does not induce a big change in our conclusion. Further we
will show in later content that a boost factor around 1.25 is within the preferred region
for detection at SK and similar Cherenkov light detectors.

The unstable particle in the final state of semi-annihilation $\psi$
is neutral under $Z_3$, and can decay to SM
particles. The decay products and decay lifetime are highly model
dependent. On the other hand, only the mass of $\psi$ can
affect the boost factor of the DM particle in the final state. Thus
focusing on the boosted DM particle provides the most
model-independent way to study this class of DM models.

We emphasize that in the simplest version of semi-annihilation
model, DM particles in the initial and final states are the same.
The scattering cross-section between DM particle and nucleon,
$\sigma_{\chi,N}$, determines both DM solar capture rate and the
interaction probability when a DM particle passes through the region
of SK. This reduces the number of effective free parameters in the model.

It is well-known that there are two classes of DM-nucleus
interactions: spin-dependent or spin-independent (SI). In traditional
DD experiments, if the scattering process is governed
by SI operators, DM can scatter with the whole nucleus
coherently due to the low velocity of DM in the local halo, leading
to strong sensitivity to DM interactions.  On the other hand, if the
scattering process is dominantly SD, the constraints
from DD are much weaker.

The situation is very different in the process we are considering.
Since the Sun is mainly constituted of hydrogen, the coherent
enhancement is absent for SI operators. There is also
a fraction of helium in the Sun, but the coherence effect is not
large enough to make order of magnitude difference on the capture
rate for SI and SD operators. On the other hand, in SK we focus
on the DM-nucleon scattering process which can kick a proton out
from the nuclei in water and generate a Cherenkov ring. The momentum transfer in
such process is generically larger than few hundreds of MeV, which
makes the coherence effects negligible. Unlike the ordinary DM
DD experiments, the reach for SI operators will be
comparable to that of SD operators. Since SD operators are much less
constrained than SI operators in DD experiments, we
choose our benchmark operators to be SD. The reach
limits for SI operators should not be dramatically different.

Furthermore, the velocity of DM particles captured and thermalized in
the Sun is much smaller than that of a boosted DM particle from DM annihilation.  In the
non-relativistic limit, $\sigma_{\chi,N}$ may have non-trivial
dependence on relative velocity or momentum transfer. In some cases,
such dependence may induce dramatic enhancements to the cross section
for interactions between boosted DM and nucleons compared to that for collisions between
the thermal non-relativistic component of DM and nuclei. In the rest of this section, we are going
to focus on two scenarios.  In the first, we assume that the leading term for
$\sigma_{\chi,N}$ in non-relativistic limit has $v^0$ dependence,
while in the second, we assume $v^2$ dependence.

\subsubsection{$v^0$ operator}
In this section, we focus on the semi-annihilating model where the
scattering cross section between DM and nucleons has $v^0$ dependence
in non-relativistic limit.  To make our discussion more concrete, we
take a benchmark operator for detailed calculations. A
typical operators of this kind is
\begin{equation}
  \label{eq:OPv0}
  \mathcal {O}_{SD,v^0}=\frac{1}{M^2}\bar{\chi}\gamma^5\gamma^\mu
  \chi \bar{q}\gamma_\mu\gamma^5 q,
\end{equation}
where $\chi$ is the DM particle, assumed to be a Dirac
fermion. $\chi_L$ and $\chi_R$ are the chiral components of $\chi$,
such that $\chi=(\chi_L,\chi_R^\dag)$. Thus $\chi_L$ and $\chi_R$ have
$Z_3$ charges as $e^{i2\pi/3}$ and $e^{i4\pi/3}$ respectively. The
scattering induced by this operator is SD.

Now let us first UV complete the effective operator in eq.\
(\ref{eq:OPv0}). This can be achieved by introducing a gauge boson
$Z'$ which couples axially to both $\chi$ and quarks. The Lagrangian
can be written as
\begin{equation}
  \label{eq:UVv0}
  \mathcal{L}\supset -i \chi_L^\dag \bar{\sigma}^\mu D_\mu \chi_L-i
\chi_R^\dag \bar{\sigma}^\mu D_\mu \chi_R- i Q^\dag \bar{\sigma}^\mu
D_\mu Q -i u^{c \dag} \bar{\sigma}^\mu D_\mu u^c-i d^{c \dag}
\bar{\sigma}^\mu D_\mu d^c
\end{equation}
We assume the axial gauge group is spontaneously broken, and $Z'$ is
massive. The covariant derivatives in eq.~(\ref{eq:UVv0}) includes
canonical couplings of $Z'$. To introduce axial current coupling, we
require $Q$ has same charge as $u^c$ and $d^c$. $\chi_L$ and
$\chi_R$ also share the same $U(1)'$ charge. The anomaly of this
$U(1)'$ can be canceled by introducing extra charged particles at
higher mass scale. For simplicity, we take the charges to have the form
$q_Q=q_{u^c}=q_{d^c}=q_{SM}$ and $q_{\chi_L}=q_{\chi_R}=q_{DM}$.
After integrating out $Z'$, the effective operator eq.~(\ref{eq:OPv0})
is generated with
\begin{equation}
  \label{eq:SuppScalev0}
  \frac{1}{M^2} =\frac{g_{Z'}^2\ q_{SM}\ q_{DM}}{m_{Z'}^2}.
\end{equation}
Noted that integrating out the longitudinal part of $Z'$ will induce
another sub-leading operator, which we drop in the following
discussion.

To place $\chi$ in the context of a semi-annihilating DM model, we
introduce another lighter Dirac fermion $\psi$, which is neutral under
$Z_3$. We take its chiral components to be $(\psi_L,\psi_R^\dag)$. Both
$\psi_L$ and $\psi_R$ have the same $U(1)'$ charge as $\chi_L$ and
$\chi_R$. The Lagrangian can be written as
\begin{equation}
  \label{eq:SAFermionLag}
  \mathcal{L} \supset
  y_1\phi\chi_L\chi_R+y_2\phi(\psi_L\psi_L+\psi_R\psi_R)+y_3\phi\psi_L\psi_R+\frac{\lambda_1}{m^2}
  (\chi_L\chi_L)(\chi_R^\dag\psi_R^\dag)+\frac{\lambda_2}{m^2}(\chi_R\chi_R)(\chi_L^\dag\psi_L^\dag),
\end{equation}
where $\phi$ is a scalar field charged under $U(1)'$. Its condensation
breaks the $U(1)'$ and gives mass to both $\chi$ and $\psi$. The
non-renormalizable terms can be easily UV completed by introducing
another complex scalar field which is also charged under $Z_3$. Its
detailed properties are not important, so we do not write down the full
UV completion explicitly. $\psi$ is not protected by $Z_3$ symmetry.
After $U(1)'$ is broken, it can decay to SM particles.

Next, we determine the cross-section for DM-nucleon interactions
using the effective operator in eq.~(\ref{eq:OPv0}). Note that the
non-relativistic scattering processes relevant to the signal
considered are in the Sun, which is made up nearly entirely of
hydrogen, which has a single proton as a nucleus, and of helium,
which has 0 spin.  Boosted DM will resolve the individual nucleons
in heavier nuclei.  Therefore, scattering off of nucleons is the
only relevant process for the process chain shown in Fig.\
\ref{fig:processes}.  The velocity of the boosted DM particle in the
final state is nearly $0.6$, so the momentum transfer cannot be
larger than roughly 2 GeV.  Therefore effective operator is always a
reasonable approximation to describe such scattering as long as the
mass of $Z'$ is larger than few GeV. Nevertheless, we present the
full form of the cross-section in terms of the phenomenological
parameters introduced below in Appendix \ref{sec:app2}.  For
simplicity, we take the approximation of $m_{\chi}\gg m_p$. If
$m_{\chi}\lesssim 5~{\rm GeV}$, most DM particles captured by the
Sun will evaporate after thermal equilibrium is reached. More
details on evaporation will be presented in later sections.

There is, however, a form factor correction relative to the proton
scattering in DD experiments, as well as possible isospin dependence
of the interactions.  For all SD interactions, the
relevant form factor is the axial form factor, which is known from
scattering neutrinos off of protons and neutrons at these energies.
The $Q^2$ dependence, as in \cite{Beacom:2003zu}, is thus
\begin{equation}
  \label{eq:formfactor}
  F(Q^2)= \frac{1}{(1+ \frac{Q^2}{M_A^2})^2},
\end{equation}
where $M_A$ is an empirical scale measured to be $1.03~\text{GeV}$,
and $\sigma_{{\rm DD}}\propto F(Q^2)^2$.

The total scattering cross-section under the above assumptions is
\begin{equation}
  \label{eq:DDv0}
  \sigma_{\chi,N}= \frac{3m_N^2m_\chi^2}{\pi M^4(m_\chi+m_N)^2}
  \left(\sum_q \Delta_q\right)^2  F(Q^2),
\end{equation}
where $\sigma_{\chi,N}$ is the scattering cross-section per
nucleon $N$, which is the same for protons and neutrons assuming
isospin-preserving interactions.  Neglecting corrections from the form
factor, if we take $\sigma_{{\rm DD},p}$ to be
$10^{-38}\textrm{cm}^2$, $M$ is around $400\ \textrm{GeV}$, where we
assumed isospin-respecting coupling to quarks, and the spin factors
$\Delta_q$ are presented in Appendix~\ref{sec:ap3}.

For a low suppression scale, one may also worry about the collider
constraints.  Leptons may be neutral under $U(1)'$, which helps in
evading constraints from LEP. Such leptophobic $Z'$ is well
motivated and has been studied extensively. Thus the only relevant
constraints are monojet and dijet resonance searches. The monojet
constraints can be alleviated by reducing the $U(1)'$ gauge coupling
or $U(1)'$ charge. To keep the effective suppression scale $M$
fixed, the mass of $Z'$ is also lowered. If the mediator mass is
smaller than twice DM mass, then the pair of DM are produced through
off-shell $Z'$ in the collider. Both smaller coupling constants and
$Z'$ being off-shell helps to avoid the monojet constraints. For
example, if we take $m_{Z'}$ to be 10 GeV while fixing the $U(1)'$
charge to be order one, to get $M=400\ \textrm{GeV}$, the $U(1)'$
gauge coupling needs to be around 0.025. With a jet cut at 250 GeV
in 8 TeV LHC, the monojet cross section is much less than 0.1 pb,
well below the current constraints from such searches. Dijet
resonance searches are very weak when $m_{Z'}$ is very light. For a
review, see \cite{Dobrescu:2013cmh}. Below 200 GeV mass (with
sensitivity down to $\sim80$ GeV), the best constraint is from UA2,
and the gauge coupling is only constrained to be smaller than 1.7 at
200 GeV. Our model is thus unconstrained from such searches as well
at the moment.

The conventional DM annihilation channel into the SM quarks must also be present due to crossing
symmetry. If this channel dominates over semi-annihilation, our signal
is suppressed and indirect detection may place strong bounds on the
model. We therefore estimate the conventional DM annihilation cross section
in our UV completion, i.e. eq.~(\ref{eq:UVv0}). Let us focus on the first generation of quarks and
take massless quark approximation. The total annihilation cross
section can be written as
\begin{equation}
  \label{eq:Annv0}
 \langle \sigma_{\rm ann}^{\chi\chi\rightarrow \bar{q}q}v \rangle= \frac{g^4 q_{SM}^2 q_{DM}^2}{2\pi} \frac{m_\chi^2 v^2}{(4m_\chi^2-m_{Z'}^2)^2+\Gamma_{Z'}^2 m_{Z'}^2}.
\end{equation}
The conventional annihilation cross section is thus p-wave suppressed. A
generic choice of parameters, this process is thus subdominant to
semi-annihilation and is unconstrained by indirect detection
experiments.

One subtlety in this model should be addressed at this point. Since we
assign axial charge to first generation quarks, the
SM Yukawa interaction is not invariant under $U(1)'$ gauge symmetry.
Setting $q_{SM}=q_{DM}$, then one can introduce a non-renormalizable
term to generate an up quark mass (with a similar term for the
generating the down quark mass),
\begin{equation}
  \label{eq:Yukawa}
  \mathcal{L}_{\rm Yukawa}= \frac{y_u }{M_{Y}} \phi\ HQ\ u^c.
\end{equation}
Small masses for first generation quarks can be generated after both
$\phi$ (which acts like a ``flavon'' here) and $H$ get VEVs. There could also be conventional DM
annihilation process through s-channel $\phi$ boson exchange. However, it is
p-wave and light quark mass suppressed.  We do not consider this
channel further.

The specific structure of this model is largely
irrelevant for the signal studied in this paper.  We thus use a
phenomenological parametrization to describe the interactions of DM
in this model. The relevant interactions are with nuclear matter.
At low energies, the cross-section for this interaction corresponds
to the DD interaction in eq.~\eqref{eq:DDv0}.  We can thus
eliminate the charges and couplings from our description in favor of
the DD cross-section, such that the parameters of the
model are taken to be the masses of $\chi$ and $Z'$, as well as the
DD cross-section $\sigma_{\rm DD}$ that would be seen
in conventional non-relativistic DD experiments.  We
emphasize once more that there is only a mild dependence on the $Z'$
mass for the parts of parameter space considered provided that
$m_{Z'}$ is larger than a few GeV.  The full differential
cross-section for DM-nucleon interactions in terms of these
parameters is presented in Appendix \ref{sec:app2} and is used in
all further calculations for such models.

\subsubsection{$v^2$ operator}
In this section, we study a semi-annihilating DM model with an
operator inducing a $v^2$ dependence in the non-relativistic
scattering between DM and nucleon. Our benchmark operator is written
as
\begin{equation}
  \label{eq:OPv2}
\frac{i}{M^2}(\chi^\dag\partial_\mu \chi -\partial_\mu \chi^\dag
\chi)\bar{q}\gamma^5\gamma^\mu q,
\end{equation}
where $\chi$ is the DM particle and is a scalar field in this model.
As with the operator in eq.~(\ref{eq:OPv0}), this is a dimension
6 operator that can be generated by integrating out a massive
gauge boson under which both SM quark and DM particle are charged.
One can UV complete this operator via the interaction Lagrangian
\begin{equation}
  \label{eq:UVv2}
\mathcal{L}\supset D_\mu \chi^\dag D^\mu\chi- i Q^\dag
\bar{\sigma}^\mu D_\mu Q -i u^{c \dag} \bar{\sigma}^\mu D_\mu u^c-i
d^{c \dag} \bar{\sigma}^\mu D_\mu d^c.
\end{equation}
We once more introduce a spontaneously broken $U(1)'$ gauge group,
with massive gauge boson $Z'$. Thus the
covariant derivatives in eq.~(\ref{eq:UVv2}) also include canonical
couplings of $Z'$. To introduce axial current coupling while forbidding
vector current coupling to quarks, we require left-handed quark $Q$ to have same charge as
$u^c$ and $d^c$. The anomaly of this $U(1)'$ can be canceled by
introducing extra charged particles at higher mass scale. For
simplicity, we take $q_Q=q_{u^c}=q_{d^c}=q_{SM}$. After integrating
out $Z'$, the effective operator eq.~\eqref{eq:OPv2} is generated with
\begin{equation}
  \label{eq:SuppScalev2}
\frac{1}{M^2}=\frac{g_{Z'}^2\ q_{\chi}\ q_{SM}}{m_{Z'}^2}.
\end{equation}
The additional contribution induced by integrating out longitudinal
component of $Z'$ is again negligible.

To add semi-annihilation processes to this scenario, we assume
scalar $\chi$ is stabilized by $Z_3$ symmetry. We introduce another
lighter scalar $\phi$, which we take to be neutral under $Z_3$. A proper
choice of $U(1)'$ charge for $\phi$ leads to the following
interaction Lagrangian:
\begin{equation}
  \label{eq:SAScalarLag}
  \mathcal{L} \supset m_{\chi}^2\chi^\dag\chi+\frac{m_\phi^2}{2}\phi^\dag\phi+\lambda_1(\chi^3 \phi^\dag+\chi^{\dag 3}\phi).
\end{equation}
It is assumed that semi-annihilation dominates over conventional
annihilation and thus determines the DM relic abundance. $\phi$ can
decay promptly after $U(1)'$ is broken. Its decay products are model
dependent and we do not study them further.

We proceed to calculate the DM-nucleon cross-section in this model.
Again, we note that in semi-annihilation scenario, it is always a
reasonable approximation to use the effective operator to calculate
the cross-section for mediator masses of at least a few GeV. $Z'$
couples axially to quarks in this model as well and one can thus
apply the same form factor as eq.~(\ref{eq:formfactor}) to estimate
the elastic scattering between DM and nucleon.

The total cross-section can be approximated by
\begin{equation}
  \label{eq:DDXsecv2}
  \sigma_{\chi,N}=\frac{m_N^2}{2\pi M^4}v^2  \left(\sum_q \Delta_q\right)^2 F(Q^2)
\end{equation}
which is suppressed by DM velocity squared, as we expect. As a
benchmark, if we take $\sigma_{\chi,N}$ to be $10^{-40}~\textrm{cm}^2$
and $v = 10^{-3}$, the suppression scale is estimated to be 30 GeV.

As with the $v^0$ operator as discussed in previous section, one
can lower $m_{Z'}$ at the same time as the coupling constant while
keeping the suppression scale $M$ fixed. For example, if we
take $m_{Z'}$ to be 5 GeV, the coupling constant should be around
0.2. Taking DM mass as 10 GeV and assuming $Z'$ couples
universally to all quarks, the monojet cross section at 8 TeV LHC
with 250 GeV jet cut is only around 0.04 pb. Further, a coupling
constant as small as 0.2 is also safe from dijet resonance
constraints. Thus collider searches are not yet sensitive to this UV
model.

As in $v^0$ scenario, one may be worried whether such low
suppression scale induces a large cross section for conventional DM
annihilation into SM quarks. To estimate the ordinary DM
annihilation cross section, we focus on the first generation of
quarks and work in the massless quark limit. The annihilation cross
section can be written as
\begin{equation}
  \label{eq:AnnXsecLowMassv2}
\langle\sigma_{\rm ann}^{\chi\chi\to q\bar{q}} v\rangle=\frac{g^4 q_{DM}^2 q_{SM}^2}{\pi}\frac{2m_{\chi}^2
v^2}{ (4m_\chi^2-m_{Z'}^2)^2+\Gamma_{Z'}^2 m_{Z'}^2}
\end{equation}
This annihilation cross section is again p-wave suppressed. A
generic choice of parameters gives a small annihilation cross
section for this standard DM annihilation channel. Thus the boosted
DM flux from the Sun will not be reduced by the existence this
channel. Also this model is safe from indirect detection
constraints.

Finally, as in the $v^0$ case, this model can be parameterized by the
phenomenological parameters $m_\chi$, $m_{Z'}$, and $\sigma_{\rm DD}$,
with only mild dependence on $m_{Z'}$.  The full differential
cross-section for DM-nucleon interactions in terms of these parameters
is presented in Appendix \ref{sec:app2} and is in future calculations.

\subsection{Two-Component DM models}
In a two-component DM model, there are at least two components of
stable particles, $\A$ and $\B$. We assume throughout that $\A$ is the
dominant component of DM. The DD constraints to $B$ can be
negligible if $B$ has a sufficiently suppressed relic abundance. In
these models, the solar capture rate is controlled by $\sigma_{A,p}$, the
scattering cross-section of $A$ with protons in the Sun,
while the boosted DM detection rate is controlled by $\sigma_{B,p}$, the
scattering cross-section of $B$ to knock out protons in the target.
As in the the semi-annihilation scenario discussed above, there
could still be subtleties coming from non-trivial velocity
dependence. However, since we have already decoupled $\sigma_{A,p}$
and $\sigma_{B,p}$, the cross section enhancement when DM is boosted
can be partially mimicked by imposing $\sigma_{B,p}\gg\sigma_{A,p}$.
Thus in the following discussion we will focus on a benchmark
model where both $\sigma_{A,p}$ and $\sigma_{B,p}$ have $v^0$
dependence at leading order.

We consider two Majorana fermion DM $\psi_A,\psi_B$, where
$m_A>m_B$. $\psi_A$ is the major DM, while $\psi_B$ is a subdominant
component. This is natural if $\psi_B$ has a larger thermal
annihilation cross section. For both $\A$ and $\B$, we consider the
same type of SD DM-nucleon scattering operator as used in
the $v^0$ dependent semi-annihilation model, i.e. eq.~(\ref{eq:OPv0}):
\begin{equation}
   \mathcal {O}_{SD,v^0}=\frac{1}{M^2}\bar{\chi}\gamma^5\gamma^\mu
  \chi \bar{q}\gamma_\mu\gamma^5 q\nonumber.
\end{equation}
Note that the fact that $\psi_A$ and $\psi_B$ are Majorana fermions
automatically eliminates the possibility of operator
$\bar{\chi}\gamma^\mu\chi$. Because of this, the quark-side of
coupling does not have to be purely axial to easily evade direct
detection constraints, as the following operator is SI
but $v^2$-suppressed \cite{Agrawal:2010fh, Cui:2010ud}:
\begin{equation}
   \mathcal {O}_{SD,v^2}=\frac{1}{M^2}\bar{\chi}\gamma^5\gamma^\mu
  \chi \bar{q}\gamma_\mu q\label{eq:OPv2_ferm}.
\end{equation}
For simplicity, we focus on the $\mathcal {O}_{SD,v^0}$ operator for
the two-component model.

Such an operator can be generated by a similar UV completion to that of
Lagrangian in eq.~(\ref{eq:UVv0}).  Both $\A, \B$ are charged under
$U(1)'$ and have Majorana masses which may result from $U(1)'$
symmetry breaking.  The relevant model Lagrangian can be written in 4-component notation as
\bea \mathcal{L}&\supset
&\bar{\psi}_A(i\partial_\mu+\frac{1}{2}q_AgZ'_\mu
)\gamma^\mu\gamma_5\psi_A+ \bar{\psi}_B(i\partial_\mu+
\frac{1}{2}q_BgZ'_\mu)\gamma^\mu\gamma_5\psi_B+\bar{\psi}_q\left[
i\partial_\mu+(g_{q,V}+g_{q,A}\gamma_5)Z'_\mu
\right]\psi_q\\\nonumber
&-&\frac{1}{2}m_A\bar{\psi}_A\psi_A-\frac{1}{2}m_B\bar{\psi}_B\psi_B-\frac{1}{2}m_{Z'}^2Z'^{\mu}Z'_{\mu},
\eea where $g$ is the $U(1)'$ gauge coupling, $q_A, ~q_B$ are the
$U(1)'$ charges of $\A,~\B$, and $\psi_q$ are the SM quarks. As
discussed earlier, for simplicity, we take $g_{q,V}\rightarrow 0$, which
can be realized in a UV construction in the same way as in the semi-annihilation models. In order
for $\B$ to annihilate away efficiently leaving a suppressed relic
abundance relative to that of $\A$, we further assume $m_B>m_{Z'}$,
which also helps alleviate the potential constraint from monojet
searches at colliders, as discussed earlier.

Both the cross sections of $\psi_A$ and $\psi_B$ for scattering off
nucleons are relevant to determining the size of the signal from
this model.  Boosted $\B$ from $\A$ annihilation can be highly
relativistic during subsequent scatterings if $m_A/m_B\gg1$.  In
cases where the $\B$ is sufficiently boosted, it is imperative to
consider the full form of the scattering cross-section for this
model as presented in Appendix \ref{sec:app2}.  Note that this model
has additional phenomenological parameters as the masses and
effective DD cross-sections for $\A$ and $\B$ are different in
general. We parameterize this model in terms of the mass and
cross-section for $\A$, as well as the ratio of the masses of the
$Z'$ and $\B$ to $\A$ and the ratio of the $\B$ effective direct
detection cross-section to that of $\A$.

In the non-relativistic limit, the total scattering cross
section with nucleons for $\chi = \A,\B$ is
 \be
\sigma^{v\to 0}_{\chi,N}=\frac{3g_\chi^2g_q^2m_\chi^2m_N^2}{\pi m_{Z'}^4(m_\chi+m_N)^2}
\left(\sum_q \Delta_q\right)^2.\ee
Various collider constraints are evaded when the mediator is light and the
couplings to quarks are not too large.

The thermal relic abundance of $\A$ is dominantly determined by
$\A\A\to\B\B$ annihilation via $A'$ exchange, with the cross section
given as follows: \be \langle\sigma_{\rm ann}^{AA\rightarrow
BB}v\rangle=\frac{g_A^2g_B^2}{12\pi}\frac{\sqrt{m_A^2-m_B^2}}{m_A(4m_A^2-m_{Z'}^2)^2+\Gamma_{Z'}^2m_{Z'}^2}
\left[3m_B^2+v^2\cdot\frac{24m_B^2m_A^4+m_A^2m_{Z'}^2(-6m_B^2+m_{Z'}^2)-m_{Z'}^4m_B^2}{m_{Z'}^4}
\right], \ee
 where we can
see that the s-wave part is suppressed by $m_B^2/m_A^2$, therefore
we have kept the potentially non-negligible p-wave contribution as
well. The relic abundance of $\A$ takes the standard form expected for WIMP DM

\be
\label{eq:relic-rel}
\Omega_A\simeq 0.2 \left( \frac{3\times10^{-26}~\text{cm}^3/\text{s}}{\langle\sigma_{A A\rightarrow B B} v\rangle} \right).
\ee

We present a couple of benchmark parameter points at which the
observed DM abundance corresponds to the thermal relic abundance of
$\A$:

{\centering $\{ m_A=150~{\rm GeV}, ~m_B=100~{\rm GeV}, ~m_{Z'}=50~{\rm GeV},
~g_A=3\cdot10^{-3}, ~g_B=0.3 \}$\\

$\{ m_A=60~{\rm GeV}, ~m_B=30~{\rm GeV}, ~m_{Z'}=20~{\rm GeV},
~g_A=0.08,~ g_B=0.4 \}$.

}

The thermal annihilation cross section of $\psi_B$ by annihilating into
$Z'$ is given by
 \be \langle\sigma_{\rm ann}^{BB\rightarrow Z'Z'}v\rangle
|_{v\rightarrow
0}=\frac{g_B^4}{2\pi}\frac{m_B^2-m_{Z'}^2}{(m_{Z'}^2-2m_B^2)^2}\sqrt{1-\frac{m_Z'^2}{m_B^2}}.
\ee
The annihilation of $\A,\B$ into SM quarks is helicity
suppressed and/or can be suppressed by assuming $g_{q}<g_{A,B}$.
Computing the thermal relic abundance of $\B$ is more complicated than
that for $\A$ as $\A\A\to \B\B$ can be important as well. Simple
analytic estimates can be obtained depending on parameter
region. These were discussed intensively in \cite{Agashe:2014yua}
where the concept of ``balanced freeze-out'' was introduced for the
region where $\B$ freezes out much later than $\A$. We do not repeat
the discussion here. Considering that there can be other channels
beyond the minimal model which can sufficiently deplete the abundance
of $\B$, in this paper we do not elaborate the relic abundance
calculation of $\B$ and related direct/indirect detection bounds.

\section{Boosted DM Flux from the Sun}
\label{sec:flux}

In this section, we determine the flux $\Phi$ of DM particles from
the Sun. The flux can be written as\footnote{It is common in the
  literature to break the
  full annihilation rate into the annihilation rate per DM pair $A$ and
  the number of DM particles $N$ as written here.}
\begin{equation}
  \label{eq:flux}
  \Phi = \frac{A N^2}{4 \pi {\rm AU}^2},
\end{equation}
where $A N^2$ is related to the annihilation rate of
DM captured in the Sun, $\Gamma_A$ by $\Gamma_A=\frac{1}{2}A N^2$, and AU is an
astronomical unit, the distance from the Sun to the Earth.  The
annihilation rate, in turn, is effectively given by the capture rate
$C$ in the part of parameter space we are interested in.  The
primary goals of this section are thus to calculate $C$ and to
determine the region of parameter space where $A N^2 = C$.

\subsection{DM Capture Rate by the Sun}
\label{sec:capture}

For any model in which DM scatters off of nucleons, the capture rate
can be written as
\begin{equation}
  \label{eq:capture}
  C = \int dV du \sigma_{\chi,H}(w \to v)\vert_{v < v_{\rm esc}} \frac{w^2}{u}
  n_\chi(r) n_H(r) f(u),
\end{equation}
where $dV$ is a volume element of the Sun, $\sigma_{\chi,H}(w \to
v)\vert_{v < v_{\rm esc}} $ is the total cross-section for DM to
scatter to a velocity below the escape velocity, $u$ is the velocity
of the DM particle if it were far away from the Sun, $w$ is the
actual velocity of the DM $\sqrt{u^2 + v_{\rm esc}^2}$, $n_\chi(r)$ is
the number density profile of DM particles near the Sun, $n_H(r)$ is the
number density of hydrogen nuclei \cite{agss09}, and $f(u)$ is the
local DM velocity distribution.  In the Appendix \ref{apps:1}, we
determine the details of the pieces of \eqref{eq:capture}.

Given these quantities, the integral over the
velocity and volume can be performed numerically.  The case where DM
has velocity suppressed interactions with matter actually has a
roughly factor of 20 enhancement of its capture rate for a fixed
effective DD cross-section.  Since the DM has fallen
into the Sun's gravitational potential, it scatters at higher
velocity in the Sun than it does in a detector near the surface of
the Earth.  In the analysis below, we determine the capture rate for
each parameter point, but in order to gain an intuition of the
orders of magnitude involved, we present the numerical results for
the parameter point with $m_\chi = 100~{\rm GeV}$ and
$\sigma_{\rm DD} = 10^{-42}~\rm cm^2$.  Here, $\chi$ is the DM
particle in semi-annihilating models and $\A$ in two component models.
The mediator mass has negligible effect since effective operator is an
excellent approximation. We thus leave it to be specified later.  In
the cases where the elastic DM-nucleon scattering cross-section is not
$v$ suppressed, we find
\begin{equation}
  \label{eq:capt-v0}
  C = 2.0 \times 10^{20}~{\rm sec}^{-1}.
\end{equation}
If the cross-section is suppressed by $v^2$, then we find
\begin{equation}
  \label{eq:capt-v2}
  C = 5.1 \times 10^{21}~{\rm sec}^{-1}.
\end{equation}
The dependence on the mass and cross-section can roughly be
parameterized in the large DM mass limit as
\begin{equation}
  \label{eq:capt-dep}
  C(m_\chi, \sigma_{\chi,N}) \approx C(100~{\rm GeV}, 10^{-42}~{\rm
    cm}^2) \left(\frac{\sigma_{\rm DD}}{10^{-42}~{\rm cm}^2}\right)
  \left(\frac{100~{\rm GeV}}{m_\chi}\right)^2
\end{equation}
for $m_\chi \gg 1~{\rm GeV}$ and for $C(100~{\rm GeV}, 10^{-42}~{\rm
  cm}^2)$ given by eq.\eqref{eq:capt-v0} or eq.\eqref{eq:capt-v2}, depending
on the model being considered.

\subsection{Capture--loss Equilibrium in the Sun}

After the formation of the Sun, DM begins to be captured by the process
described in Sec.~\ref{sec:capture}.  On the other hand, there are
two dominant processes that reduce the amount of DM in the Sun,
annihilation and evaporation. After a long time of accumulation of DM
particles in the Sun, an equilibrium state may be reached such that
\begin{equation}
  \label{eq:semi-ann-master}
  A N^2 = C - E N,
\end{equation}
where $E$ is the rate per DM particle at which DM evaporates from
the Sun and $N$ is the number of DM particles captured in the Sun.
In a regime where $E \approx 0$, which we demonstrate below is
generic, the annihilation rate is given by
the capture rate. After DM particle is captured by the Sun, it will
soon reach thermal equilibrium with the Sun.  The DM distribution can be
characterized by a thermal radius, which is given by \cite{Nussinov:2009ft}
\begin{equation}
  \label{eq:ThermalRadius}
r_{\rm th}=\left(\frac{3T}{2\pi m_\chi G_N \rho_c}\right)^{1/2}=0.01
R_{\odot}\left(\frac{T}{1.2\ \textrm{keV}}\right)^{1/2}\left(\frac{100\
\textrm{GeV}}{m_\chi}\right)^{1/2},
\end{equation}
where $G_N$ is Newton's constant, $\rho_c$ is the core density of the
Sun, $R_{\odot}$ is the solar radius, and $T=1.2\rm keV$ is the
Sun's core temperature. Thus one can calculate the time needed for the DM to reach an
equilibrium between capture and annihilation, $\tau_{\rm eq}=1/\sqrt{C\cdot A}$, and compare it with
the age of the Sun \cite{Nussinov:2009ft}:
\begin{equation}
  \label{eq:Equilibrium}
\frac{t_\odot}{\tau_{\rm eq}}=10^3\left(\frac{C}{10^{25}\
\textrm{sec}^{-1}}\right)^{1/2}\left(\frac{\langle\sigma_{\rm ann}v\rangle}{3\times
10^{-26}\ \textrm{cm}^3 \textrm{sec}^{-1}}\right)^{1/2}\left(\frac{0.01\
R_\odot}{r_{\rm th}}\right)^{3/2}.
\end{equation}

As long as $\frac{t_\odot}{\tau_{\rm eq}}>1$, equilibrium is
reached by the present day. In this case, the DM flux will be independent of
the DM  annihilation cross section.  For an annihilation cross-section
close to the value which gives correct thermal relic abundance, the
Sun will have reached capture--annihilation equilibrium by now. Even
in models where the relic abundance is determined by p-wave
annihilation, as long as there is a non-negligible s-wave component,
equilibrium can still be reached in the Sun. For the models we will be
discussing, the equilibrium condition will always be satisfied for the
region of interest in parameter space.

A full treatment of DM evaporation in the case without velocity
suppression is found in \cite{Gould:1987ju}.  Here, we apply simple
arguments to estimate the evaporation rate in the $v^2$ case.
Evaporation occurs when an energetic nucleus on the tail of the
(local) solar Boltzmann distribution collides with a slower DM
particle and imparts a velocity larger than the escape velocity.
Typically, the DM velocity will be much less than the nucleus
velocity and we approximate it to be at rest.  Under this
approximation, the evaporation rate per unit volume in the Sun is
given by
\begin{equation}
  \label{eq:evap}
  \frac{dE}{dV} = \int_{(m_p + m_\chi) v_{\rm esc}/2 m_p}^\infty du \sigma_{\chi,H}(0 \to v)\vert_{v > v_{\rm esc}} n_\chi(r) n_H(r)
  u_H f(u_H),
\end{equation}
where $\mu$ is the proton-DM reduced mass, $\sigma(0 \to v)\vert_{v
>
  v_{\rm esc}}$ is the cross-section for a proton of velocity $u_H$
to impart a velocity above the escape velocity, $n_\chi$ is the
local captured DM number density, $n_H$ the local hydrogen density,
and $f$ is the velocity distribution of hydrogen in the Sun.  The
individual pieces are again presented in the Appendix.\ref{apps:1}.  The
resulting rate can be integrated.  It is found that in this
approximation, the term $E N$ in \eqref{eq:semi-ann-master} can be
neglected for DM masses above $4~{\rm GeV}$.  To be conservative
given the approximations we make in this determination, we consider
only DM masses larger than 5 GeV and neglect evaporation effects
entirely in models with either $v^0$ and $v^2$ behavior.

\subsection{Rescattering in the Sun}
\label{sec:rescattering-sun}

As the DM particle travels from the center of the Sun where it is
produced in annihilation processes, it may rescatter off of solar
nucleons and lose velocity.  This alters the effective detection
cross-section.  In this paper, we use a conservative estimate for
this effect.  At larger cross-sections, there is a larger flux and thus a larger
detection rate for boosted DM.  On the other hand, there is an
increasingly significant loss of energy as the DM exits the Sun.
Since only DM with sufficiently large energy can scatter protons to
momenta above the Cherenkov threshold, most of the scattered DM will
not be detected if the energy loss is too great.  Since both the
collision rate with the detector target and the energy loss as the
DM exits the Sun scale as $\sigma_{\rm DD}^2$, there is a detailed
interplay between these effects.  In this paper, we use
the very conservative approximation that there is no sensitivity to
models for which the mean energy of the DM escaped from the Sun is
insufficient to provide a large enough Cherenkov detection rate. For
larger cross-sections, fluctuations become important in determining
the mean detection cross-section and there is likely sensitivity up
to significantly higher cross-sections.  A full determination of the
mean detection cross-section is beyond the scope of this work.

The easiest way to compute this mean energy at the exit of the Sun is
to work with Mandelstam variables.  The probability that a DM particle
interacts with Mandelstam variable $t$ between $t$ and $t + \Delta t$
while traveling between $r$ and $r + dr$ from the center of the Sun is
given by
\begin{equation}
  \label{eq:inter-prob}
  dP = n_N(r) \frac{d\sigma}{dt}(s,t) dt dr,
\end{equation}
where $n_N$ is the number density of nucleons in the Sun\footnote{In
  this calculation, we assume isospin is a valid approximation such
  that the scattering cross-sections off protons and neutrons are equal.}.
 Recall that in the rest frame of the nucleons,
 $s=m_{\chi}^2+m_N^2+2E_\chi m_N$, $t=2m_N(E'_{\chi}-E_{\chi})$.
Therefore the change in energy is proportional to the change in Mandelstam
$s$, which is given by
\begin{equation}
  \label{eq:change-sigma}
  ds = t dP,
\end{equation}
where Mandelstam $t$ is the change in $s$ for a collision
parameterized by $s$ and $t$.  The mean energy is directly related
to the mean value of $s$.  To determine this mean, we solve
\begin{equation}
  \label{eq:loss-eff-sigma}
  \int_{s_0}^{\langle s \rangle} \frac{ds}{\int
    dt ~t (d\sigma/dt)} = \int_0^{R_{\odot}} dr n_N,
\end{equation}
where $s_0 \equiv s(v_0)$, $v_0$ is the constant velocity of boosted DM
particles coming right out of the annihilation process, and $R_\odot$ is the
solar radius.  We do not claim to have sensitivity to models where
the detection rate assuming DM particles incident on the detector with
energy $\langle E_\chi \rangle$, which is directly related to $\langle s
\rangle$, is too small to generate a sufficient number of signal events.

\section{Detection of Boosted DM}
\label{sec:detect}

\subsection{Detection Mechanism for Signals}
Neutrino experiments have well established techniques to detect
recoils of energetic charged particles and thus can be repurposed to
detect boosted DM particles via recoiling protons or electrons from
DM-matter collision. The flux of boosted DM from the Sun
is small compared to, say, that of the non-relativistic halo DM,
so that a large volume detector is required. Two representatives of
the largest active neutrino experiments (including their near future
upgrade/extension) are IceCube/Deepcore/PINGU and SK/HK, which both use
photo-multiplier tubes to detect Cherenkov light emitted during collisions with
the target, and are potentially good candidates for boosted DM
search. In this work, we focus on detecting Cherenkov protons instead
of electrons from DM-matter collisions in the detector for the
following reasons. First, as discussed earlier, for neutral-current
type of interaction, scattering off protons has a larger rate than off
electrons, except for models where very $t$-channel light mediator is
involved \cite{Agashe:2014yua}. Second, since the DM solar
capture and rescatter rates are determined by the interaction between DM and
nucleon, focusing on detecting proton signal at terrestrial
experiments avoids introducing further model dependence in the
DM-electron coupling.

The detection of proton recoils with momentum larger than around
$2~{\rm GeV}$ becomes problematic.  In this regime, scattering
becomes dominantly inelastic, leading to multi-rings events where
the direction information is lost. Additionally, protons above
$2~{\rm  GeV}$ in momentum have $>50\%$ chance of producing pions as
they travel through the detector, whose decay would give an extra
electron-like ring. Finally, high energy elastic collisions may be
more difficult to distinguish from muon recoils.  As illustrated in
\cite{Fechner:2009aa}, a Cherenkov ring of a recoiling proton is
similar to that of a muon, but different from that of an electron
which has blurred edge due to electromagnetic showers. Proton with
sufficiently low momenta, less than a few GeV, are likely to be
stopped within the detector due to its strong nuclear interactions,
which causes the Cherenkov light emission from the recoiling proton
to stop abruptly.

Due to the high energy threshold of IceCube, $\gtrsim100$ GeV, it is
not suitable to detect proton tracks in the a few GeV range. It is
expected to be challenging even at its low energy extension, PINGU,
with a few GeV as energy threshold\cite{priv_greg}.
In this paper we focus on determining the current
limits and expected sensitivity for SK and HK in the single-ring
proton track channel.  Planned Liquid Argon Time-Projection Chamber
experiments \cite{Bueno:2007um,Badertscher:2010sy} may also have
sensitivity to proton recoils, though a study of this prospect is
beyond the scope of this work.

For the reasons outlined above, when studying detection at SK and HK,
we consider only recoiling proton momenta below $2~{\rm GeV}$ as in
\cite{Beacom:2003zu}. Since the target material is water,
which has a proton Cherenkov momentum of $1.07~{\rm GeV}$
\cite{Beacom:2003zu}, there is only sensitivity for collisions which
yield proton momenta at least this energetic.

Given the detection mechanism of searching for single Cherenkov
rings from protons recoiling in water, we proceed to determine
the effective cross-section of the detector and thereby obtain a
estimated prediction for the number of expected signal events.  This
effective cross-section is determined both by the detection
efficiency and acceptance and by the short distance scattering of
the DM off of the target protons.

The typical recoil spectrum for protons above the Cherenkov
threshold in the regions of parameter space considered in this work
are similar to those of the background neutrinos.  Given this
similarity, we estimate the efficiency for detecting a DM particle
to be given by $70\%$, which is the estimated efficiency for detecting
atmospheric neutrinos via their single proton Cherenkov
signature\cite{Fechner:2009aa}.

\subsection{Background Reduction}
The main background for our signal is from atmospheric neutrinos with
NC interaction, which are nearly
isotropic across the sky, and are the aim of the current searches for proton tracks at SK.
Additional backgrounds to detection include those that fake
atmospheric neutrino neutral current scattering.  As outlined in
\cite{Fechner:2009aa}, these are primarily charged pion and muon scattering
events from cosmic rays. Our signal can be distinguished from these
backgrounds based on the following discriminators:
\bi
\item \textit{Angular information:}

Whereas the atmospheric neutrino background is nearly isotropic, the
incoming boosted DM is coming nearly entirely from the direction of
the Sun.  The signal may be enhanced compared
to the atmospheric neutrino and other backgrounds by cutting on the
angle of the recoiling proton with respect to the Sun.  For a
boosted DM velocity of $v = 0.6$, assuming $m_{DM}\gg m_p$, the
maximal angle between the incoming DM particle and the recoiled
proton is $40.9^\circ$ from the Sun.  The angular resolution for the
recoiling protons is $2.8^\circ$ which is not an effective limiting
factor in this case. We find that optimal $s/\sqrt{b}$ is obtained
for a cut on the proton recoil angle equal to the maximal recoil
angle, $\theta_{\rm max} = 40.9^\circ$ for $v = 0.6$, around the
Sun.  This cut is optimized for the semi-annihilation DM spectrum,
but is also approximately correct for two-component models in the
region of parameter space to which our signal provides the greatest
sensitivity. Beyond the requirement that the proton momentum fall
between 1.07 GeV and 2 GeV, the signal acceptance for this cut is
essentially 1. The background acceptance, on the other hand, is
reduced by roughly ratio of the solid angle covered by the search to
$4\pi$,
\begin{equation}
  \label{eq:bkg-accept}
  \eta_{\rm bkg} \approx \frac{1}{2} (1 - \cos\theta_{\rm max}) = 0.122.
\end{equation}

In our analysis, we use the most recent SK proton track data that is
publicly available, which includes the runs of SK-I and SK-II, up to
the year of 2009 \cite{Fechner:2009aa}. With this data set, the
background for our search is expected to be 49.6 in a lifetime of
2287.8 days over the full detector acceptance\cite{Fechner:2009aa}.
After applying the above acceptance cut, we would expect 6.05
background events.

In addition to eliminating a significant fraction of the background,
an angular restriction would leave a large solid angle of side band,
which would have no signal contamination.  This region could be used
to normalize the background and determine the detection efficiency,
thereby eliminating most of the systematic uncertainty in the
measurement of the signal region.  For this reason, for our projected
sensitivity calculations, we do not consider any systematic
uncertainty, assuming that systematics would be subdominant to
statistical uncertainties.
\item \textit{Absence of $\mu^\pm, e^\pm$ excesses:}

Neutrino background giving rise to proton tracks via NC interaction
also lead to corresponding electron and muon signals via
charged-current (CC) interaction with comparable rate. But those
accompanying channels are absent for boosted DM as it only has NC
type of interactions (as discussed the $e^-$ signal from DM-$e^-$ NC
scattering can in generally suppressed, and has a different
correlation to the proton signals compared to the case of
neutrinos). This feature can only distinguish boosted DM signal from
possibly proton tracks generated by the neutrinos from conventional
WIMP DM annihilation in the Sun, which cannot be reduced by the
directional information we discussed earlier.

\ei

Further background discrimination may be possible by exploiting
finer information such as the energy or angular distribution of the
scattered proton or by using multi-variate analysis. Here we
take a basic cut-and-count approach based on simple variables.

We now turn to determining the short distance
contribution to the effective detector cross-section. Given a DM
particle that has exited the Sun with velocity $v$, corresponding to
a Mandelstam $s$ in collisions with protons at rest,  the effective
cross-section for the DM to be detected by the detector can be
written as
\begin{equation}
  \label{eq:eff-sigma-def}
  \Sigma(s) = \epsilon(s) Z \sigma_{\rm Cher}(s),
\end{equation}
where $\epsilon$ is the detection efficiency, $Z$ is the number of
protons in the detector and $\sigma_{\rm Cher}$ is the
cross-section for a DM particle to scatter a proton within the
accepted momentum range as described above. Since the typical
momentum transfer is very high, the  proton binding to other
nucleons is negligible.  As discussed above, we take $\epsilon(s)$,
which does in principle depend on the incident DM velocity, to be a
constant $70\%$. For SK, we have $Z = 6.8 \times 10^{33}$, while for
HK, the planned target has $Z = 1.7 \times 10^{35}$.  The threshold $s$ for
Cherenkov detection is
\begin{equation}
  \label{eq:cer-thresh}
  s_{\rm min} = \sqrt{m_p (2 m_\chi^2 E_{p,{\rm Cher}}+ 2 m_p
    m_\chi^2 + m_p p_{{\rm Cher}^2)}} + m_p E_{p,{\rm Cher}} + m_\chi^2.
\end{equation}

\section{Results}\label{sec:results}
The flux and effective cross-section calculated in
Sections.\ref{sec:flux} and \ref{sec:detect} can be combined to approximate the
total number of events expected in the detector as
\begin{equation}
  \label{eq:total-evts}
  N_{\rm sig} = \Phi \Sigma(\langle s \rangle) \Delta t,
\end{equation}
where $\Delta t$ is the lifetime of the experiment. We conduct the
following three different analyses to evaluate the sensitivity of SK
and HK for the boosted DM search as we proposed.

In the first analysis, using the $CL_s$ method we determine the 95\%
exclusion region implied by the current SK analysis published in
\cite{Fechner:2009aa}, which combines runs I and II of the SK
experiment with lifetime 2287.8 days and assumes signal coming from
the full solid angle of the sky.  The number of proton recoil events
observed in SK runs I and II is 38, with an expected atmospheric
neutrino and other process background totaling 49.6 events
\cite{Fechner:2009aa}.  The background is assumed to follow a
log-normal distribution centered at 49.6 events with a 20\% width,
which accounts for estimated systematic and theory uncertainties. We
marginalize over the background distribution.  Given the data
observation and expected background, as well as a signal efficiency
of $70\%$, models which predict more than $20.7$ events before
folding in the detection efficiency are currently excluded.

The second analysis determines the expected sensitivity if the
analysis were extended to include runs III and IV of SK, as well as
making use of the direction of proton recoils relative to the Sun,
as described in Section.\ref{sec:detect}.  Finally, we determine the
expected sensitivity of the HK experiment assuming a lifetime equal
to that of runs I-IV of the SK experiment and again making use of
the proton recoil direction. The lifetime of runs I-IV is 4438.2
days. In these two analyses we determine $2\sigma$ sensitivity
region assuming the signal is present.  The expected backgrounds for
the full SK and HK datasets, after applying a cut on the recoil
angle with respect to the Sun, are $23.5$ and $587$ events
respectively. Note that since we are focusing on a small angular
region for these analyses, we expect systematic uncertainties to be
greatly reduced by looking at a side-band away from the direction of
the Sun, such that statistic uncertainty dominates for background
estimation. We find that a $2\sigma$ excess can be claimed if $15.5$
events and $230$ events respectively are predicted before folding in
the detection efficiency.

\begin{figure}
  \centering
  \includegraphics[width=0.49\textwidth]{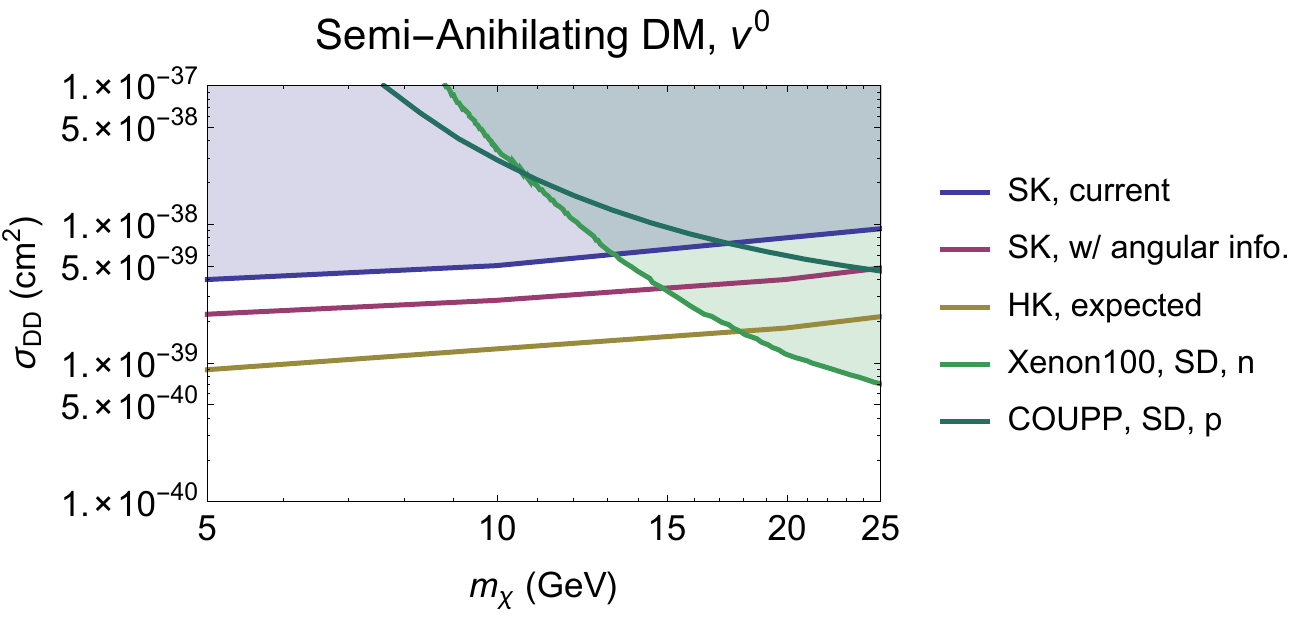}
  \includegraphics[width=0.49\textwidth]{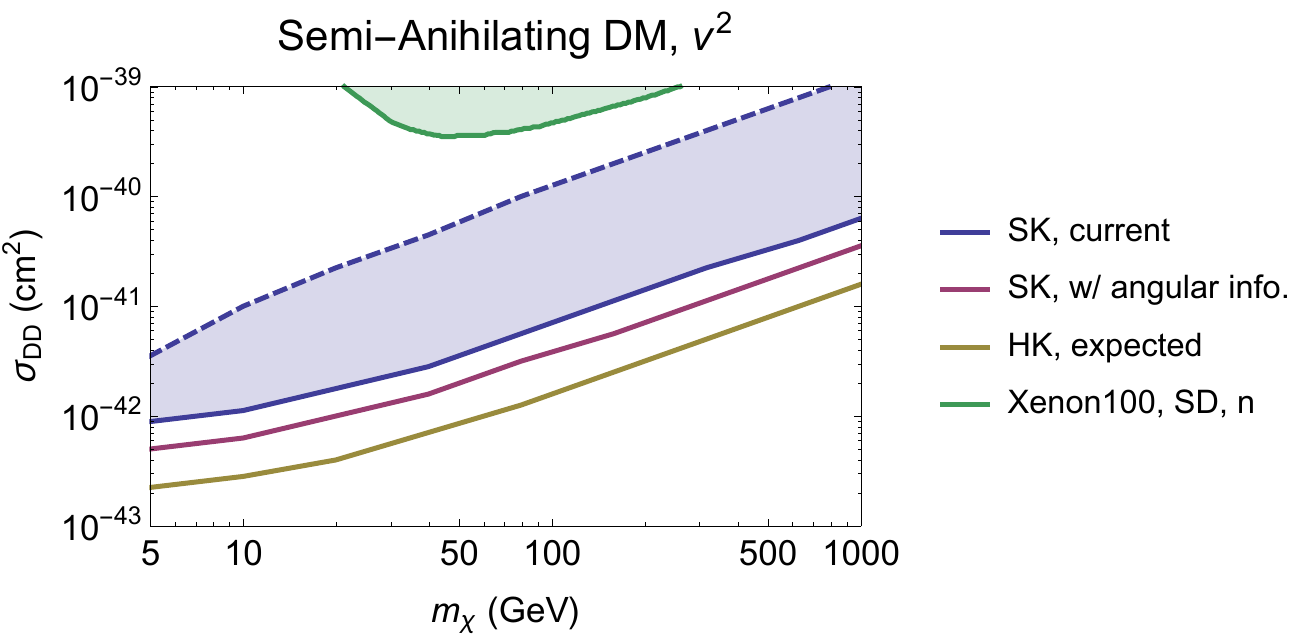}
  \caption{Limits on the parameter space of semi-annihilation models
    with low energy operators $O_{SD,v^0}$ (left) and $O_{SD,v^2}$
    (right).  The lines labeled as ``SK, current,'' ``SK, w/ angular info.,'' and
    ``HK, expected'' are obtained as described in \ref{sec:results}.   The
    dashed line right above the ``SK, current'' line indicates the cross-section above
    which rescattering lowers the mean velocity such that the
    detection cross-section at the mean velocity is too low to be
    seen.  This is not a hard cutoff, but rather a conservative
    estimate, as described in the Sec.~\ref{sec:detect}.  The lines labeled as
    ``Xenon100, SD, n'' and ``COUPP, SD, p'' are
    derived from Refs. \cite{Aprile:2013doa,coupp}.  The models are
    parameterized by their effective DD cross-section and
    the DM mass.}
    \label{fig:sa-results}
\end{figure}

\begin{figure}
  \centering
  \includegraphics[width=0.49\textwidth]{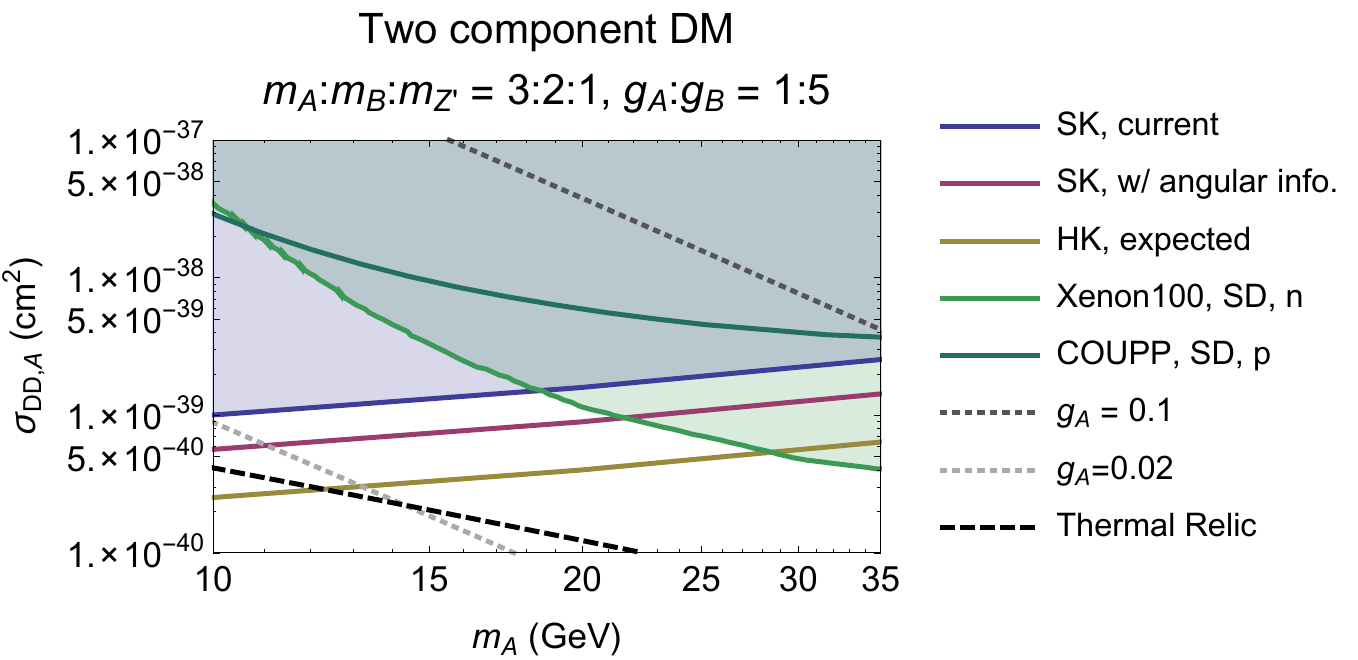}
  \includegraphics[width=0.49\textwidth]{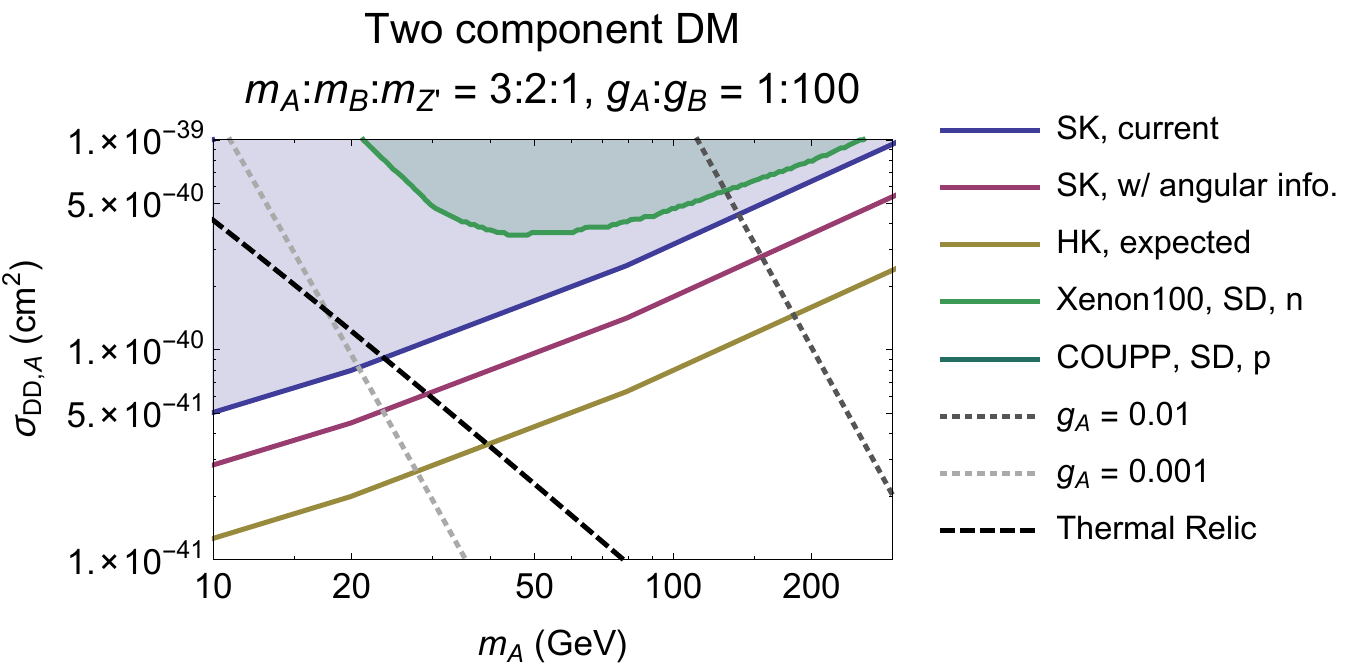}
  \caption{Limits on the parameter space of two-component models for two
    different benchmark parameter choices.  The lines labeled as ``SK,
    current,'' ``SK, w/ angular info.,'' and ``HK, expected'' are
    obtained as described in \ref{sec:results}. The 
    lines labeled ``Xenon100, SD, n'' and ``COUPP, SD, p'' are derived from
    Refs. \cite{Aprile:2013doa,coupp}.  The models are parameterized by the
    effective DD cross-section and the DM mass for the
    heavier DM particle $\A$, which is assumed to make up nearly $100\% $
    of cosmic DM.  The thermal relic lines assume $\Omega_A = 0.23$
    and are derived using eq. \ref{eq:relic-rel}.  For
    reference, lines of corresponding to models with fixed coupling
    $g_A$ are shown as well.}
    \label{fig:hier-results}
\end{figure}
From the above limits, we determine the model parameter space for which $N_{\rm sig}$
events are excluded by current SK data and for which there would be
sensitivity using the full data at SK or HK and applying a cut on
the recoil angle of the proton. The results for the
semi-annihilating models are presented in Fig.\
\ref{fig:sa-results}.  It is emphasized once more that these results
are essentially independent of the specific structure of the model,
beyond the Lorentz structure of the DM-quark scattering amplitude.
The results for two benchmark two-component DM models are presented
in Fig.~\ref{fig:hier-results}.  While these results do introduce
some model dependence via the ratio of couplings and masses, rough
scaling of the limits holds as
\begin{equation}
  \label{eq:scaling-hier}
  \sigma_{{\rm DD},A}^{\rm lim}(m_B/m_A,g_B/g_A) \sim \sigma_{{\rm DD},A}^{\rm
    lim}(2/3,5)  \left(\frac{g_B/g_A}{5}\right)^2,
\end{equation}
provided that $1.1 \lesssim m_A/m_B \lesssim 2.2$.  Beyond this mass
range, \eqref{eq:scaling-hier} may not be a good approximation, but,
particularly for smaller mass ratios, there should still be a
significant signal.  Such models would predict even more energetic
incident DM particles in the detector and the resulting proton
collisions would be beyond the regime where the approximations we
make hold.  A full Monte Carlo simulation study including inelastic
scattering is likely required to determine the constraints.  Note
that a larger $g_A/g_B$ ratio tends to give more viable, reachable
parameter space. Even for the large choice of $g_A/g_B=100$ as in
the right plot of Fig.~\ref{fig:hier-results}, $g_B$ is still well
within perturbative region. Nonetheless such large coupling
hierarchy may be unnatural if both $\A$ and $\B$ couples to the same
$Z'$ directly. One may resolve this by assuming $\A$ couples to another
$U(1)'$ gauge boson which mixes with the $Z'$ with which $\B$ couples
directly. Alternatively, $\A$, $\B$ may couple to quarks
through two separate $Z'$s entirely, with $\A\A\to\B\B$ generated by
an separate operator. These all require going beyond the minimal model,
yet are still reasonable possibilities. Lastly, we emphasize that
the reason we propose the large ratio of $g_B/g_A$ is to induce a
sizable enhancement of $\sigma_{{\rm DD},B}$ respect to $\sigma_{{\rm DD},A}$.
This can also be achieved by assuming $\A$-proton and $\B$-proton
scattering are induced by different effective operators. Thus the
a large ratio between $\sigma_{{\rm DD},B}$ and $\sigma_{{\rm DD},A}$ can
be easily achieved in other scenarios.

\section{Conclusion}\label{sec:conclusions}

In this paper, we investigate a new search channel for boosted DM.  As
opposed to earlier related work \cite{Agashe:2014yua}, which focused on ``secluded'' DM where observable signals are only possible from DM annihilation in the galactic center, here we consider here
the alternative case where the DM also has appreciable interactions
with SM quarks so that it can be effectively captured in the Sun, where it
annihilates and produces boosted DM that can be detected at large
volume neutrino detectors. Such
annihilation also determines the thermal relic abundance of the DM,
which realizes thermal WIMP paradigm in an alternative way. We study both semi-annihilating DM and two-component DM as examples.
 In our simple example models, this
detection mechanism provides additional sensitivity beyond
conventional direct detection and collider searches in models with
relatively light mediators that couple to light quarks and dominantly
generate spin-dependent DM-nucleus scattering.  Viable alternative models are
possible and may be worth investigating.

We propose a new search based on elastic scattering induced proton
tracks pointing towards the direction of the Sun, which is typically
the primary search channel for DM and mediator with weak scale
masses. In particular, we studied the sensitivity limits for boosted
DM at Cherenkov-light based neutrino detector such as SK and its
upgrade HK. The existing SK proton track data already has sensitivity
to some of the parameter space, while the region that could be probed
would be moderately enlarged by using an analysis that takes into
account directional information with the full SK data and would be
even more enlarged by a search at HK.  Future large-volume liquid
Argon neutrino detector based on ionization signals may significantly
extend the sensitivity. As already studied in
\cite{Agashe:2014yua}, the single ring electron channel may be
complementary to or even more important than the proton track channel
when a mediator mass much lower than weak scale is involved. It is
also possible to have complementary indirect detection signals, for
instance from $\phi$ decay in semi-annihilation model with
$\chi\chi\to\chi^\dag\phi$ or $Z'$ decay in $\B\B\to Z'Z'$ in
two-component model. We leave an investigation of these interesting possibilities to future work.

Boosted DM is generic in scenarios with multiple DM
components or single DM specie with non-minimal stabilization
symmetry. Detecting boosted DM can be crucial in uncovering such features
of DM sector, and in some cases can even be a smoking gun signal. As
we have seen, even in the models where appreciable DM-nucleon
scattering rate is present to facilitate solar capture of DM, current
or near future DD limit can still be far from the 
parameter region that can be probed by dedicated boosted DM search. It is
particularly intriguing that large-volume neutrino detectors can be
repurposed to search for boosted DM.  These signals provide further
motivation for a thorough examination of new search strategies for
DM sector with non-minimal structure.

\section*{Acknowledgement}
We thank Kaustubh Agashe, Valentin Hirschi, Junwu Huang, Ed Kearns, Lina Necib, Greg Sullivan and
Jesse Thaler for discussions.  SLAC is operated by Stanford University
for the US Department of Energy under contract DE-AC02-76SF00515.  YC
is supported by Perimeter
Institute for Theoretical Physics, which is supported by the
Government of Canada through Industry Canada and by the Province of
Ontario through the Ministry of Research and Innovation. YC is in
part supported by NSF grant PHY-0968854 and by the Maryland Center
for Fundamental Physics. YZ is supported by ERC grant BSMOXFORD no.
228169.
\appendix

\section{Isospin Dependence of Spin-Dependent DM-nucleon Scattering}\label{sec:ap3}
The isospin dependence of DM-nucleon scattering can be parameterized
as
\begin{equation}
  \label{eq:32}
  \sigma_{\chi,N} = \sigma_{\chi p} \frac{[A_u \Delta_u^{(n)} + A_d
    \Delta_d^{(n)} + A_s \Delta_s^{(n)}]^2}{[A_u \Delta_u^{(p)} + A_d
    \Delta_d^{(p)} + A_s \Delta_s^{(p)}]^2},
\end{equation}
where $A_{u,d,s}$ parameterize the relative sizes of the couplings
of DM to the quark flavors and $\Delta_{u,d,s}^{(N)}$ are given by
\cite{Belanger:2008sj}
\begin{equation}
  \label{eq:33}
  \Delta_u^{(p)} = 0.78, \qquad \Delta_d^{(p)} = -0.48, \qquad \Delta_S^{(p)}
  = -0.15
\end{equation}
and
\begin{equation}
  \label{eq:33}
  \Delta_u^{(n)} = \Delta_d^{(p)}, \qquad \Delta_d^{(n)} = \Delta_u^{(p)},\qquad \Delta_s^{(n)}
  = \Delta_s^{(p)}.
\end{equation}
Note that if the coupling to quarks is universal, then this factor is
1.  For simplicity and concreteness, we assume this case.

\section{Parametrization of the DM-nucleon Cross-section}\label{sec:app2}

Beyond the discussion in Section \ref{sec:models}, we parametrize all
of the relevant cross-sections using the effective DD
cross-section $\sigma_{\rm DD}$, rather than the couplings.  This
parameter is related to the Lagrangian parameters by
\begin{equation}
  \label{eq:rel-d-v0}
  \sigma_{\rm DD} = \frac{3 \mu_{\chi,N}^2}{\pi M^4} \left(\sum_q \Delta_q\right)^2,
\end{equation}
for models corresponding to the operator in eq.~\eqref{eq:OPv0} with
$\chi$ denoting the DM particle in semi-annihilating models and $\chi
= \A,\B$ in two component models.  $M$ is the suppression scale of the
corresponding operator.  For the model corresponding to eq.\
\eqref{eq:OPv2}, the relation is
\begin{equation}
  \label{eq:rel-dd-v2}
  \sigma_{\rm DD} = \frac{m_N^2}{2\pi M^4}v_0^2  \left(\sum_q \Delta_q\right)^2,
\end{equation}
where $v_0^2$ is the mean squared velocity of DM in the local halo.

In terms of these cross-section ``parameters,'' the full form of the
scattering cross-section for DM-nucleon interactions for $v^0$-like
interactions is \be \frac{d\sigma_{\chi,N}}{dt}=\frac{\sigma_{\rm
DD}}{24
\mu_{\chi,N}^2}\cdot\frac{m_{Z'}^4}{(t-m_{Z'}^2)^2}\frac{t^2+2t(2E_\chi
m_p-m_p^2-m_\chi^2)+8m_p^2(E_\chi^2+2m_\chi^2)}{\lambda(s,m_p^2,m_\chi^2)}F(-t)^2,
\ee where $E_\chi$ is the energy of incoming DM $\chi$ in the lab frame,
and in the lab frame $t=2m_p(m_p-E_p)$, $s=m_\chi^2+m_p^2+2E_\chi m_p$,
$E_p$ is the energy of the scattered proton, $\lambda(x,y,z)\equiv
x^2+y^2+z^2-2xy-2yz-2zx$, and $F$ is the form factor given in eq.\
(\ref{eq:formfactor}).  For $v^2$-like interactions, the full form
of the differential cross-section is
\begin{equation}
  \frac{d\sigma_{\chi,N}}{dt} = - \frac{\sigma_{\rm DD}}{8 \mu_{\chi,N}^2
    v_0^2} \frac{m_{Z'}^4}{(t-m_{Z'}^2)^2} \frac{t (m_\chi +
    m_N)^2}{m_N^2 p_\chi^2} F(-t),
\end{equation}
where $p_\chi$ is the lab-frame momentum of $\chi$.

\section{Detailed Determination of the Capture and Evaporation Rates}\label{apps:1}

In this Appendix, we calculate the rates $C$ and $E$ presented in
\eqref{eq:capture} and \eqref{eq:evap} respectively.  We proceed
factor by factor in the integrands.

A DM particle $\chi$ that has velocity $w$ at solar radius $r$ and
scatters off of a nucleus $n$ in the Sun is
captured if its final velocity $v$ in the solar frame is less than the
escape velocity $v_{\rm esc}$ at the radius of the scattering.  This
cross-section is given by
\begin{equation}
  \label{eq:capture-xs}
  \sigma_{\chi,p}(w \to v)|_{v < v_{\rm esc}} =
  \int_{-1}^{c_{\theta,{\rm max}}} dc_\theta \frac{d\sigma_{\chi,p}}{dc_\theta},
\end{equation}
where $c_\theta$ is the cosine of the CM scattering angle.  The
minimum scattering angle is given by
\begin{equation}
  \label{eq:28}
  c_{\theta,{\rm max}} = 1 - \frac{m_n m_\chi}{p_{\chi,{\rm CM}}^2}
  \left(\frac{1}{\sqrt{1 - v_{\rm esc}^2}} - \frac{1}{\sqrt{1 - w^2}} \right),
\end{equation}
where $p_{\chi,{\rm CM}}$ is the momentum of the DM particle in the CM
frame.
At the low momentum transfers in the collisions in the Sun, the DM
scatters coherently off of the nuclei in the Sun.  Since the
scattering is SD in the models we consider and helium is
spin 0, scattering occurs exclusively on hydrogen nuclei, which are
protons. The escape velocity is determined from solar model
\cite{agss09}.

The DM velocity going into these collisions is determined from the
local DM velocity distribution.  The velocity at distances far from
the Sun is approximately given by a Boltzmann distribution
\cite{Gould:1987ir}
\begin{equation}
  \label{eq:boltzmann}
  f(u) = \sqrt{\frac{6}{\pi}} \frac{u}{v_G \bar{v}} \exp\left(-\frac{3}{2}
    \frac{u^2 + v_G^2}{\bar{v}^2}\right) {\rm sinh} \left(\frac{3 u
      v_G}{\bar{v}^2}\right),
\end{equation}
where $v_G$ is the velocity of the Sun in the Milky Way and $\bar{v}^2$
is the local mean squared velocity of DM.  As the DM falls into the
gravitational potential of the Sun, it gains speed such that, by
conservation of energy,
\begin{equation}
  \label{eq:velocity-collision}
  w(r) = \sqrt{u^2 + v_{\rm esc}^2(r)}
\end{equation}
at distance $r$ from the center of the Sun.  In addition, the density
of DM in the Sun gets a Sommerfeld enhancement from falling into the
Sun of $w/u$.

The local number density of DM far from the Sun is taken to be $m_\chi
n_\chi = 0.3~{\rm GeV}/cm^3$.  The number density of hydrogen atoms at
a distance $r$ from the center of the Sun is again determined from the
solar model \cite{agss09}.

Putting all of these pieces together, we find that the rate for DM
capture in a volume element $dV$ near a distance $r$ from the center
of the Sun and coming from velocity between $u$ and $u + du$ at
distance far from the Sun is given by
\begin{equation}
  \label{eq:capture-v2}
  dC = dV du (\sigma_{\chi,H}(w \to v)|_{v < v_{\rm esc}} w n_H)
  \left(\frac{w}{u} n_\chi \right) f(u),
\end{equation}
where the first factor is the number of interactions at $r$ per DM
particle and the second factor is the number density of DM including
the Sommerfeld enhancement.  To obtain \eqref{eq:capture}, we
integrate over the velocities $u$ for which scattering to $v <
v_{\rm esc}$ is possible and over the volume of the Sun.  The upper
limit on $u$ such that capture is possible at a radius $r$ is given
by
\begin{equation}
  \label{eq:upper-u}
  u < \frac{2 \sqrt{m_\chi m_p} v_{\rm esc}}{m_\chi - m_p}.
\end{equation}
for $m_\chi > m_p = m_H$.

The determination of the evaporation follows similar arguments.  The
primary differences are in the detailed kinematics.  For instance, the
minimum scattering angle is given by
\begin{equation}
  \label{eq:evap-max}
  c_{\theta,{\rm max}} = 1 + \frac{m_\chi^2}{p_{\chi,{\rm CM}}^2} \left(1 -
    \frac{1}{\sqrt{1 - v_{\rm esc}^2}}\right).
\end{equation}
There is no Sommerfeld enhancement, but the DM density is given by the
captured DM density.  If the DM undergoes a sufficient number of
scatters before it annihilates, then its number density is thermalized
and given by
\begin{equation}
  \label{eq:thermal-dm-dist}
  n_\chi(r) = N \frac{\exp\left(- M \Phi(r) / T_W
    \right)}{\int_0^{R_\odot} dr 4 \pi r^2 \exp\left(- M \Phi(r) / T_W
    \right)},
\end{equation}
where $\Phi(r)$ is the gravitational potential of the Sun at $r$ and
$T_W$ is the thermalized DM temperature.  The thermalized DM
temperature is an averaged solar temperature, which we take to be
$T(r)$, the temperature of the Sun at $r$.

The distribution of hydrogen velocities follows a thermal distribution
at $r$ given by
\begin{equation}
  \label{eq:dist-hydro}
  f(u_H) = \left(\frac{m_p T(r)}{2 \pi}\right)^{3/2} \exp\left(-
    \frac{m_p u_H^2 }{2 T(r)}\right).
\end{equation}
In order to induce evaporation, the hydrogen velocity must be
sufficiently large to kick the DM to a velocity above the escape
velocity.  In the non-relativistic limit, this minimal velocity is
given by
\begin{equation}
  \label{eq:min-uh}
  u_H > \frac{(m_p + m_\chi) v_{\rm esc}}{2 m_p}.
\end{equation}

\end{document}